\documentclass[a4paper]{article}
\usepackage{ASVspoof2021}
\usepackage{epsfig,amssymb,amsmath,url}
\usepackage{xcolor}
\usepackage{subcaption}
\usepackage{multirow}
\usepackage{balance}
\usepackage{verbatim}
\ninept

\title{ASVspoof 2021: accelerating progress in \\
spoofed and deepfake speech detection}

\makeatletter
\def\name#1{\gdef\@name{#1\\}}
\makeatother
\name{
      {\em Junichi Yamagishi, Xin Wang, Massimiliano Todisco, Md Sahidullah, Jose Patino,}\\
      {\em Andreas Nautsch, Xuechen Liu, Kong Aik Lee,  Tomi Kinnunen, Nicholas Evans, Héctor Delgado}\\
}      

\address{ASVspoof consortium\\
{\tt \small \url{http://www.asvspoof.org/}}}
\begin{document}
\maketitle

\begin{abstract}

ASVspoof 2021 is the fourth edition in the series of bi-annual challenges which aim to promote the study of spoofing and the design of countermeasures to protect automatic speaker verification systems from manipulation.  In addition to a continued focus upon logical and physical access tasks in which there are a number of advances compared to previous editions, ASVspoof 2021 introduces a new task involving deepfake speech detection.  This paper describes all three tasks, the new databases for each of them, the evaluation metrics, four challenge baselines, the evaluation platform and a summary of challenge results.  Despite the introduction of channel and compression variability which compound the difficulty, results for the logical access and deepfake tasks are close to those from previous ASVspoof editions.  Results for the physical access task show the difficulty in detecting attacks in real, variable physical spaces.  With ASVspoof 2021 being the first edition for which participants were not provided with any matched training or development data and with this reflecting real conditions in which the nature of spoofed and deepfake speech can never be predicated with confidence, the results are extremely encouraging and demonstrate the  substantial progress made in the field in recent years.
\end{abstract}

\section{Introduction}

The ASVspoof initiative \cite{wu2017asvspoof,9358099} was conceived to foster progress in the development of countermeasures (CM) to protect automatic speaker verification (ASV) systems from spoofing attacks.  ASVspoof 2021 is the forth edition in a series of biannual challenges \cite{Wu-ASVspoof2015,Kinnunen2017-assessing,asvspoof2019} in which participants are tasked with the design of CMs for one or more tasks.
The 2021 edition \cite{ASVspoof21_evalplan} includes three such tasks: logical access (LA); physical access (PA); speech deepfake (DF).  The LA task builds upon the ASVspoof 2015 \cite{Wu-ASVspoof2015} and ASVspoof 2019 LA tasks \cite{asvspoof2019}.  With a telephony scenario in mind, the LA task involves the detection of synthetic and converted speech when these are considered to have been injected into a communication system (e.g.\ a telephone line) without any acoustic propagation.  The 2019 LA task built upon the first 2015 edition by considering state-of-the-art neural and acoustic waveform models.  The 2021 edition extends the challenge further by considering telephony encoding and transmission.

The 2021 PA task returns from the simulation of replay attacks to a setup closer to that of the ASVspoof 2017 challenge \cite{Delgado2018}.  For the 2021 edition, both bona fide access attempts and source replay data are recorded in a variety of real physical spaces according to a strictly controlled setup using a number of different loudspeakers, recording and replay devices. Replay attacks are then played and re-recorded under the same conditions.  In contrast to the LA task, the scenario includes acoustic propagation.  All audio data hence contains reverberation and additive noise.

The DF task is new to ASVspoof and extends the focus of the initiative to the detection of spoofed speech in non-ASV scenarios.  The DF task reflects the scenario in which an attacker has access to the voice data of a targeted victim, e.g.\ data posted to social media.  The victim might be a celebrity, a social media influencer, or an ordinary citizen. 
The attacker's incentive might be, e.g., to blackmail the victim or to denigrate his or her reputation in some way by spreading \emph{spoken misinformation}.  The attacker is assumed to use the public data and speech deepfake technology to generate spoofed speech resembling the voice of the victim, and then to post the recordings to social media, call centers, or any other such application.  

We present an overview of the different challenge tasks, the databases designed for each, and the chosen metrics.  While the LA and PA tasks retain use of the tandem detection cost function (t-DCF) \cite{Kinnunen-tDCF2020}, with no ASV system, we revert to the equal error rate (EER) for the DF task.  As in previous years, ASVspoof 2021 adopts a number of different baseline solutions that are also described in the paper.  Finally, we report a summary of system performance for the baseline systems and those submitted by ASVspoof 2021 challenge participants.

\section{Database}

The ASVspoof 2021 database comprises three new evaluation partitions for the LA\footnote{https://doi.org/10.5281/zenodo.4837263}, PA\footnote{https://doi.org/10.5281/zenodo.4834716}, and DF\footnote{https://doi.org/10.5281/zenodo.4835108} tasks. 
With no new training or development data being released, participants were expected to use the training and development partitions of the ASVspoof 2019 databases~\cite{asvspoof2019database} based upon speech derived from the VCTK base corpus\footnote{http://dx.doi.org/10.7488/ds/1994}. 
These contain speech recording collected from 20 speakers (8 male, 12 female) for the training partitions and 10 speakers (4 male, 6 female) for the development partitions.  The new LA and PA evaluation partitions comprise speech recordings collected from the same 48 speakers (21 male, 27 female) corresponding to the ASVspoof 2019 evaluation partition.  Data for the DF task is derived not only from the VCTK base corpus, but other (undisclosed) corpora as well. 

ASVspoof 2021 is intentionally more challenging than previous editions: the 2021 LA evaluation data contains new trials for each speaker and both encoding and transmission artefacts introduced by real telephony systems; the new PA evaluation data is based upon the same source data as the ASVspoof 2019 data, but is recorded in real, variable physical spaces in which there is additive noise and reverberation; the new DF evaluation data exhibits audio coding and compression artefacts and also includes data captured in different domains.  
Countermeasures must therefore generalise well to a multitude of different nuisance variation as might be expected in many practical applications.
Despite the numerous sources of variation, all evaluation data for all tasks was distributed to ASVspoof participants in a common 16-bit PCM format with a 16~kHz sampling rate, compressed in FLAC format.
Presented in the following is a brief summary of 
each of the three evaluation datasets.

\subsection{Logical Access}

The ASVspoof 2021 LA evaluation data includes a collection of bona fide and spoofed utterances transmitted over a variety of telephony systems including voice-over-IP (VoIP) and a public switched telephone network (PSTN).  The transmission of all data across telephony systems introduces nuisance variability that might be expected in a range of real logical access application scenarios.  While no noise is added to the recordings, the data may reflect artefacts introduced not only as a result of spoofing, but also as a result of encoding and transmission.

\begin{table}[]
    \centering
    \caption{Summary of LA data conditions}
    \begin{tabular}{|c|c|c|c|}
        \hline
        Cond. & Codec	& \begin{tabular}{@{}c@{}} Audio \\ bandwidth \\ \end{tabular}   & Transmission \\ 
        \hline\hline
        LA-C1 & - & 16 kHz & - \\ 
        LA-C2 & a-law & 8 kHz & VoIP \\ 
        LA-C3 & unk. + $\mu$-law & 8 kHz & PSTN + VoIP \\ 
        LA-C4 & G.722 & 16 kHz & VoIP \\ 
        LA-C5 & TBA & 8 kHz & VoIP \\ 
        LA-C6 & TBA & 8 kHz & VoIP\\ 
        LA-C7 & TBA & 16 kHz & VoIP \\ 
        \hline
    \end{tabular}
    \label{tab:LA_conditions}
\end{table}

While spoofed trials result from one of 13 different speech synthesis and voice conversion spoofing attack algorithms (A07-A19 conditions in~\cite{asvspoof2019database}), for all but one condition, both spoofed and bonafide trial data was treated with one of seven distinct codecs as a result of transmission, giving the seven evaluation conditions listed in Table~\ref{tab:LA_conditions}.  Condition C1 replicates the ASVspoof 2019 LA scenario, i.e.\ with neither transmission nor encoding.  
Conditions C2 and C4-C7 correspond to transmission across an Asterisk\footnote{\url{https://www.asterisk.org}} private branch exchange (PBX) system using one of five different codecs operating at either 8~kHz or 16~kHz bandwidths (e.g.\ a-law and G.722).   C3 relates to the transmission over a PSTN system starting from a mobile smartphone and ending at a SIP endpoint hosted on a professional VoIP system which uses a $\mu$-law codec at an 8~kHz bandwidth.  This particular condition may hence reflect the application of multiple, unknown intermediate PSTN transcodings and network transmissions.  With the post-evaluation phase still active at the time of writing, the specific codecs used for conditions C5-C7 remain undisclosed for the time being.  They will be revealed during the ASVspoof 2021 workshop.

Last, the distribution in speakers and spoofing attacks
is balanced in each condition meaning that differences in detection performance for each condition can be attributed to differences in encoding and transmission.  Additional metadata which will allow for more detailed analyses will be released in due course.

\subsection{Physical Access}

\begin{figure}[t]
\centering
\includegraphics[trim=0 330 0 80, clip, width=\linewidth]{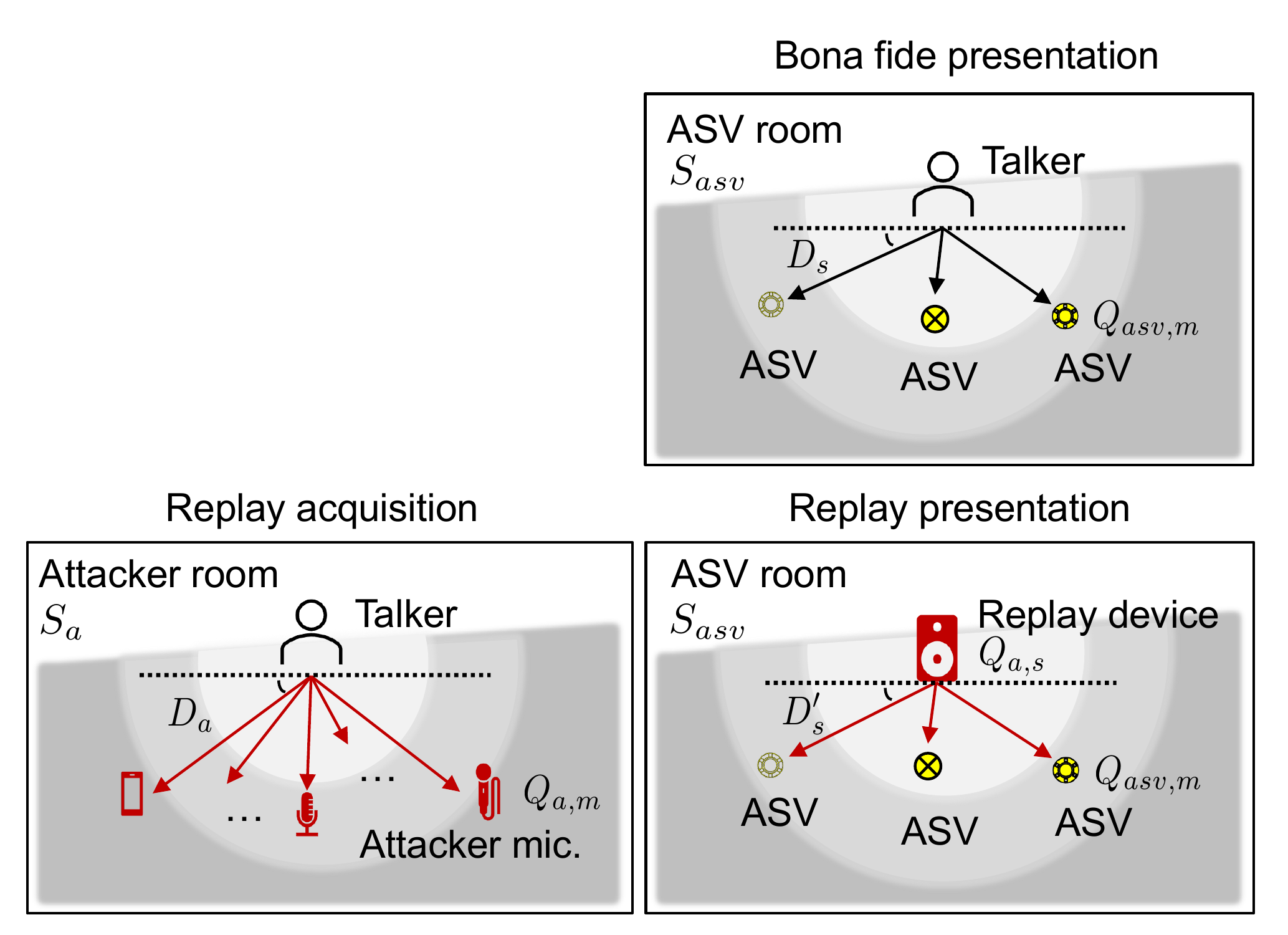}
\vspace{-5mm}
\caption{An illustration of the ASVspoof 2021 physical access task. Rooms for replay acquisition and presentation may be different. The talker is simulated by a high quality loudspeaker. 
}
\label{fig:pa-recording}
\end{figure}

The ASVspoof 2021 PA evaluation data comprises real bona fide and replayed samples similar to the ASVspoof 2017 database, but with a better controlled design similar to that of the ASVspoof 2019 PA database. 
Bona fide trials are presented to the ASV system in a real physical space whereas replay attacks are acquired and then re-presented using devices of varying quality. The recording processes are illustrated in Fig.~\ref{fig:pa-recording}.  Accordingly, PA evaluation data contains reverberation and additive noise from both the environment and replay devices. 

Bona fide data is seeded from the VCTK corpus.
Seed utterances are presented to the ASV system using a high-quality loudspeaker that has a reasonably flat frequency response (rather than being uttered by a genuine talker\footnote{We adopt the term \emph{talker} instead of \emph{speaker} to avoid confusion with the loudspeaker device used in mounting replay attacks.}). 
Recordings are made in nine different rooms listed in Table \ref{tab:PA_conditions} in which are situated three different types of microphones
at each of six different talker-to-ASV distances.  Recordings using a total of 18 microphones from each are made simultaneously. There are hence 162 ($=9 \times 3 \times 6$) different evaluation environments.

\begin{table}[t]
    \centering
    \caption{Summary of PA evaluation conditions. The meaning of each factor is illustrated in Fig.~\ref{fig:pa-recording}.}
    \resizebox{\columnwidth}{!}{%
    \begin{tabular}{|c|c|c| c |c|c|c|}
        \cline{1-3}
         & \multirow{2}{*}{Cond.}   & Room size & \multicolumn{4}{c}{} \\ 
        \cline{5-7}
         &  & $w\times d \times h (m)$ &  & & Cond. & Angle, Dis.(m)  \\ 
        \cline{1-3}\cline{5-7}
        \multirow{9}{1.2cm}{$S_{\text{asv}}$ / $S_{\text{a}}$} & R1 / r1 & $8.0\times8.0\times2.4$ &  & \multirow{6}{1.cm}{$D_{s}$, $D_{s}'$}    & D1 & $15^{\circ}$,  2.0  \\ 
                                                             & R2 / r2 & $6.0\times5.0\times2.3$ &  &                                           & D2 & $45^{\circ}$,  1.5  \\ 
                                                             & R3 / r3 & $6.6\times5.0\times2.4$ &  &                                           & D3 & $75^{\circ}$,  1.0  \\
                                                             & R4 / r4 & $7.5\times7.7\times2.6$ &  &                                           & D4 & $90^{\circ}$,  0.5  \\
                                                             & R5 / r5 & $7.2\times4.0\times2.3$ &  &                                           & D5 & $120^{\circ}$, 1.25 \\
                                                             & R6 / r6 & $4.5\times6.5\times2.5$ &  &                                           & D6 & $150^{\circ}$, 1.75 \\
        \cline{5-7}
                                                             & R7 / r7 & $4.5\times2.4\times2.4$ &  & \multirow{3}{*}{$D_{a}$}   & d1 & TBA  \\ %
                                                             & R8 / r8 & $7.1\times4.8\times2.5$ &  &                            & d2 & TBA  \\ %
                                                             & R9 / r9 & $5.9\times4.0\times2.8$ &  &                            & d3 & TBA  \\ %
        \cline{1-3} \cline{5-7}
        \multicolumn{7}{c}{} \\ 
        \cline{1-3} \cline{5-7}
        &  Cond. &  Mic. type &  & & Cond. & Device type \\ 
        \cline{1-3} \cline{5-7}
        \multirow{3}{1.2cm}{$Q_{\text{asv,m}}$} & M1 & MPM-1000 (condenser)  & & \multirow{3}{*}{$Q_{a, s}$}  & s2 & TBA \\
                                                                    & M2 & Uber Mic (condenser) & &  & s3 & TBA \\
                                                                    & M3  & iPad Air (MEMS)  & &  & s4 & TBA \\
        \cline{1-3}\cline{5-7}
        \multirow{3}{1.2cm}{$Q_{\text{a,m}}$} & m1 &  TBA & \multicolumn{4}{c}{} \\
                                                                    &  m2 &  TBA & \multicolumn{4}{c}{} \\
                                                                    &  m3 &  TBA & \multicolumn{4}{c}{} \\
        \cline{1-3} 
    \end{tabular}
    }
    \label{tab:PA_conditions}
\end{table}

Replays are made in the same rooms used to record bona fide trials. Spoofed trials for the replay are recorded both in that and two other rooms.
The set of different attacker factors include the room size, attacker microphone device, attacker-to-talker distance, attacker replay device, and the attacker-to-ASV distance~\cite{asvspoof2019database}.
With the number of exhaustive attack factor combinations being large, we adopt a non-exhaustive policy whereby replay attacks in any one room are made using one of nine different attack factor combinations and they are re-recorded with the 18 ASV microphones.
Specific attacker replay devices, microphones and attacker-to-talker distance used for the spoofed trials remain undisclosed for the time being. They will be revealed during the ASVspoof 2021 workshop.
Both bona fide and spoofed trials are downsampled to 16~kHz and pre-processed to remove leading and ending silence segments exceeding 0.3 seconds.

\subsection{DeepFake}

The DF evaluation data is a collection of bonafide and spoofed speech utterances processed with different lossy codecs used typically for media storage. Audio data is encoded and then decoded to recover uncompressed audio. This process introduces distortions that depend on the codec type and configuration. Evaluation source data is taken from the ASVspoof 2019 LA evaluation set,
as well as from other sources, resulting in 2021 evaluation data comprising spoofing attacks generated with more than 100 different 
spoofing algorithms.

As listed in Table~\ref{tab:DF_conditions}, three different codecs were used in generating the DF evaluation data, together with an additional `no codec' condition C1.  Condition C2 and C3 both use an mp3 codec. C4 and C5 use an m4a code, whereas both C6 and C7 use an ogg codec.  The differences between conditions with the same codec lie in use of different variable bit rate (VBR) configurations, one low bit rate setting and one high bit rate setting.  The specific bitrate range in each case is listed in the last column of Table~\ref{tab:DF_conditions}. Two additional conditions C8 and C9 use other, non-disclosed compression techniques, thereby giving a total of nine different compression conditions.  We used ffmpeg\footnote{\url{http://ffmpeg.org/}} and sox\footnote{\url{http://sox.sourceforge.net/}} toolkits in creating the DF evaluation data. 

With the post-evaluation phase still ongoing at the time of writing, codec condition C8 and C9 will be revealed at the ASVspoof 2021 workshop, as will the additional data sources.

\begin{table}[]
    \centering
    \caption{Summary of DF evaluation conditions. Each of these conditions also includes different vocoder types as sub-conditions.}
    \begin{tabular}{|c|c|c|}
        \hline
        Cond. & Compression	& VBR (kbps)    \\ 
        \hline\hline
        DF-C1 & - & -  \\ 
        DF-C2 & mp3 & $\sim$80-120  \\ 
        DF-C3 & mp3 & $\sim$220-260  \\ 
        DF-C4 & m4a & $\sim$20-32 \\ 
        DF-C5 & m4a & $\sim$96-112 \\ 
        DF-C6 & ogg & $\sim$80-96\\ 
        DF-C7 & ogg & $\sim$256-320 \\ 
        DF-C8 & TBA &  TBA \\
        DF-C9 & TBA &  TBA \\
        \hline
    \end{tabular}
    \label{tab:DF_conditions}
\end{table}

\section{Performance measures}

ASVspoof 2021 adopts two different performance metrics. The primary metric for the LA and the PA tasks is the so-called \emph{tandem detection cost function (t-DCF)} \cite{Kinnunen-tDCF2020}. It assesses the combined (tandem) performance of CMs and ASV, viewing the CM as a `bonafide/spoof gate' placed before an unprotected ASV system. The metric reflects the Bayes' risk and can be adjusted for different applications by choosing different detection cost and class prior parameters. 

Note that the t-DCF metric presented in \cite{Kinnunen2018-tDCF} (and used in ASVspoof 2019) was simplified in \cite{Kinnunen-tDCF2020} to include fewer parameters. While lengthy derivations and discussions are available in \cite{Kinnunen-tDCF2020,ASVspoof21_evalplan}, the metric used in ASVspoof 2021 has a simple form, 
    \begin{equation}\label{eq:tdcf}
        \text{t-DCF} = \min_{\tau}\Big\{C_0 + C_1\, P_\text{miss}(\tau) + C_2\, P_\text{fa}(\tau)\Big\},
    \end{equation}
where $P_\text{miss}(\tau)$ and $P_\text{fa}(\tau)$ are miss and false alarm rates of a CM as a function of detection threshold $\tau$, while $C_0$, $C_1$ and $C_2$ are cost function parameters. These parameters depend not only on pre-defined cost and prior parameters but on ASV performance as well; $C_0$ and $C_1$ depend on ASV performance for bonafide trials while $C_2$ increases linearly with the ASV system spoof false alarm rate.  Hence, the more detrimental the spoofing attack is to the ASV system, the higher the penalty for CM false alarms.

The parameters $C_0$, $C_1$ and $C_2$ are different for the LA and PA tasks (see Table \ref{tab:tdcf-coeffs}), and for their respective progress and evaluation partitions. While a `moving target' metric may appear strange (typical challenge metrics remain fixed across progress/evaluation data), this reflects the tandem operation of the two systems: the design objective of the CM varies according to ASV performance.

\begin{table}[!t]
    \centering
    \addtolength{\tabcolsep}{-2pt}
    \caption{t-DCF parameters for the LA and PA tasks.}
    \label{tab:tdcf-coeffs}
    \begin{tabular}{|c|ccc|ccc|}
        \hline
        & \multicolumn{3}{c|}{progress} & \multicolumn{3}{c|}{evaluation}\\
        Task & $C_0$ & $C_1$ & $C_2$ & $C_0$ & $C_1$ & $C_2$\\
        \hline\hline
         LA & 0.1588 & 2.1007 & 0.8412 & 0.1847 & 
         2.0173 & 
         0.8153\\
         PA & 
         0.1363 & 1.6345 & 0.8637 & 
         0.1291 & 
         1.6800 & 
         0.8709 \\
         \hline
    \end{tabular}
    
\end{table}

As in ASVspoof 2019, the ASV system is set to operate at its target-nontarget EER threshold, obtained from data pooled across all bonafide segments in a given task (LA or PA) and data partition (progress or evaluation). Different to ASVspoof 2019, where the $C_0$ term in \eqref{eq:tdcf} --- the \emph{ASV floor} --- was discarded, ASVspoof 2021 retains $C_0$. As seen from \eqref{eq:tdcf}, the ASV floor is the t-DCF obtained with an error-free CM (no misses or false alarms). It reflects errors caused by the ASV system.

The primary metric for the DF task is the \emph{equal error rate} (EER). It reflects the capability of a given CM to discriminate between bonafide and spoof utterances. Since the DF task does not include an ASV system, we opted for the general EER metric that does not require specification of cost and prior parameters. 
The EER serves as an upper bound to the best achievable (aka Bayes) error rate (e.g.~\cite{brummer2021trials}).

\section{ASV system and CM baselines}

\subsection{ASV system}
The ASV system comprises a DNN-based \emph{x-vector}~\cite{Snyder2018XVectorsRD} speaker embedding extractor, along with a backend based on \emph{probabilistic linear discriminant analysis} (PLDA)~\cite{prince2007probabilistic}.  The embedding extractor is trained using 30-dimensional MFCC features using data collected from 6114 speakers contained in the VoxCeleb2~\cite{voxceleb2} database.  The embedding extractor was implemented with the open-source ASVTorch toolkit~\cite{asvtorch}. The x-vector model consists of five frame-level \emph{time-delay neural network} (TDNN) layers and statistics pooling that maps the frame-level representations to utterance-level representations.%
Two 512-dimensional fully-connected layers and a softmax output then follow. Embeddings are extracted from the first fully-connected layer, without activation and normalization.

We trained a Kaldi~\cite{povey2011kaldi} PLDA-based backend using the VoxCeleb1 database~\cite{nagrani2017voxceleb}. 
Adaptation was applied separately for the LA, PA and DF tasks using the techniques described in~\cite{plda_adaptation}. Adaptation was performed using the bonafide speech utterances in the ASVspoof 2019 CM training partitions which were processed with codecs (LA), simulated acoustic propagation effects (PA), or compression algorithms (DF).  The scaling factor of the within-class and between-class covariances 
are tuned 
to different tasks: $\alpha_{w}=0.25, \alpha_{b}=0.0$ for LA and DF; $\alpha_{w}=0.9, \alpha_{b}=0.0$ for PA.  All speaker embeddings for PLDA training and adaptation were centered, unit length normalized, and transformed with linear discriminant analysis (LDA).

\begin{table}[!t]
    \centering
    \begin{tabular}{|c|c|cc|}
    \hline
         Scenario & Condition & Prog & Eval \\ \hline
         \multirow{2}{1.75em}{LA} & \emph{Bonafide} & 6.38 & 7.62 \\ \cline{2-4}
         & \emph{Spoofed} & 36.84 & 38.50 \\ \cline{1-4}
         \multirow{2}{1.75em}{PA} & \emph{Bonafide} & 6.99 & 6.48 \\\cline{2-4}
         & \emph{Spoofed} & 48.37 & 45.44 \\ \cline{1-4}
         \multirow{2}{1.75em}{DF} & \emph{Bonafide} & 3.24 & 2.79 \\\cline{2-4}
         & \emph{Spoofed} & 35.22 & 42.06\\ \cline{1-4}
    \end{tabular}
    \caption{ASV pooled preliminary baseline EER(\%) results on different scenarios. \emph{Spoofed} results are pooled across different spoofing conditions. Prog and Eval correspond to progress and evaluation sets, respectively.}
    \label{tab:asv_results}
\end{table}

ASV EERs for each task and for progress and evaluation partitions of the ASVspoof 2021 database are illustrated in Table~\ref{tab:asv_results}.  The differences in EERs for bonafide and spoofed conditions serve to show the vulnerability of the ASV system to spoofing; EERs for the spoofed condition are universally higher than for the bonafide condition.

\subsection{CM baselines}
Four different CM baseline systems were used for ASVspoof 2021.
The first two are GMM-based systems which operate on either constant-Q cepstral coefficients (CQCCs)~\cite{TODISCO2017516} (baseline B1) or linear frequency cepstral coefficients (LFCCs)~\cite{Sahidullah15} (baseline B2). 
CQCC features are extracted using 12 bins per octave. Re-sampling is applied with a sampling period of 16.  Features comprise 19 static cepstra plus energy, delta and delta-delta coefficients. LFCC features are extracted using a 30~ms window with a 15~ms shift, a 1024-point Fourier transform and 70 filters.  Features comprise 19 static cepstra plus energy, delta and delta-delta coefficients.
Both LFCC-GMM and CQCC-GMM systems are applied to data with a maximum frequency of 4~kHz for LA and DF tasks and 8~kHz for the PA task. 

The third LFCC-LCNN system~\cite{wang2021cm} (baseline B3) operates upon LFCC features with a light convolutional neural network (LCNN).  Features are similar to the LFCC-GMM but the frame length and shift are set to 20ms and 10ms respectively. The back end is based on the LCNN reported in~\cite{lavrentyeva2019stc}, but incorporates LSTM layers and average pooling.

The forth system~\cite{tak2021end} (baseline B4) uses a RawNet2 architecture~\cite{Jung2020}.  As a fully end-to-end system,  it operates directly upon raw audio waveforms.
It consists of a fixed bank of sinc filters, 6 residual blocks followed by a gated recurrent units (GRU) and fully connected (FC) layers prior to the output layer. A softmax activation function is applied in order to produce two-class predictions: bona fide or spoof.

All baseline systems were trained using only the respective ASVspoof 2019 training data and optimised using only the respective development data (DF systems used LA data).  None used any kind of data augmentation.

\begin{figure}[t]
\centering
\includegraphics[width=0.45\textwidth]{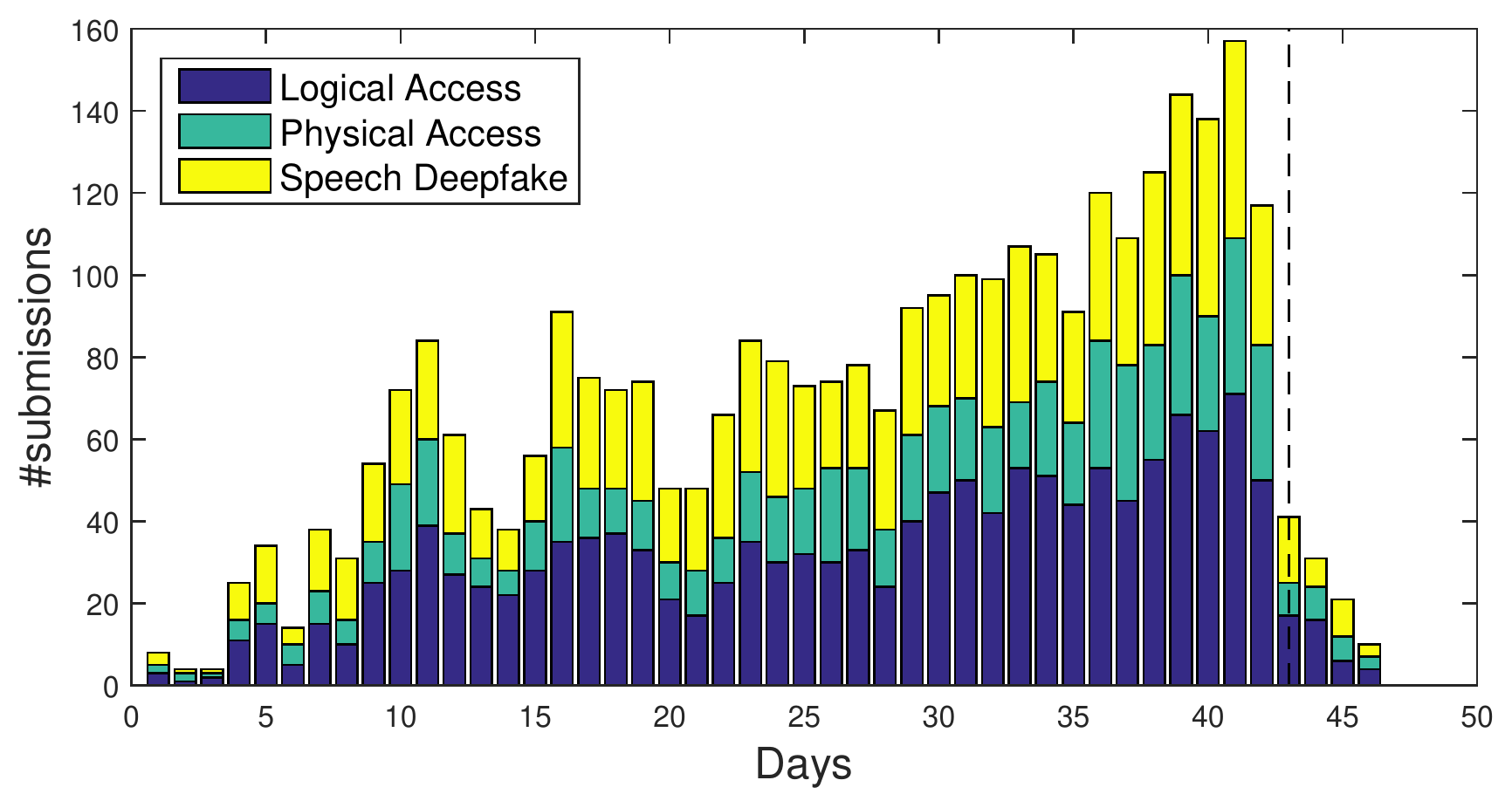}
\vspace{-3mm}
\caption{A \emph{stacked} bar chart showing the number of submissions to the LA, PA, and DF tasks for the \emph{progress} and \emph{evaluation} phases, where the latter started from day 43 onward (transition marked with vertical dashed bar).}
\vspace{-3mm}
\label{fig:la_pa_df_progress_eval_count}
\end{figure}

\begin{figure}[t]
\centering
\includegraphics[width=0.45\textwidth]{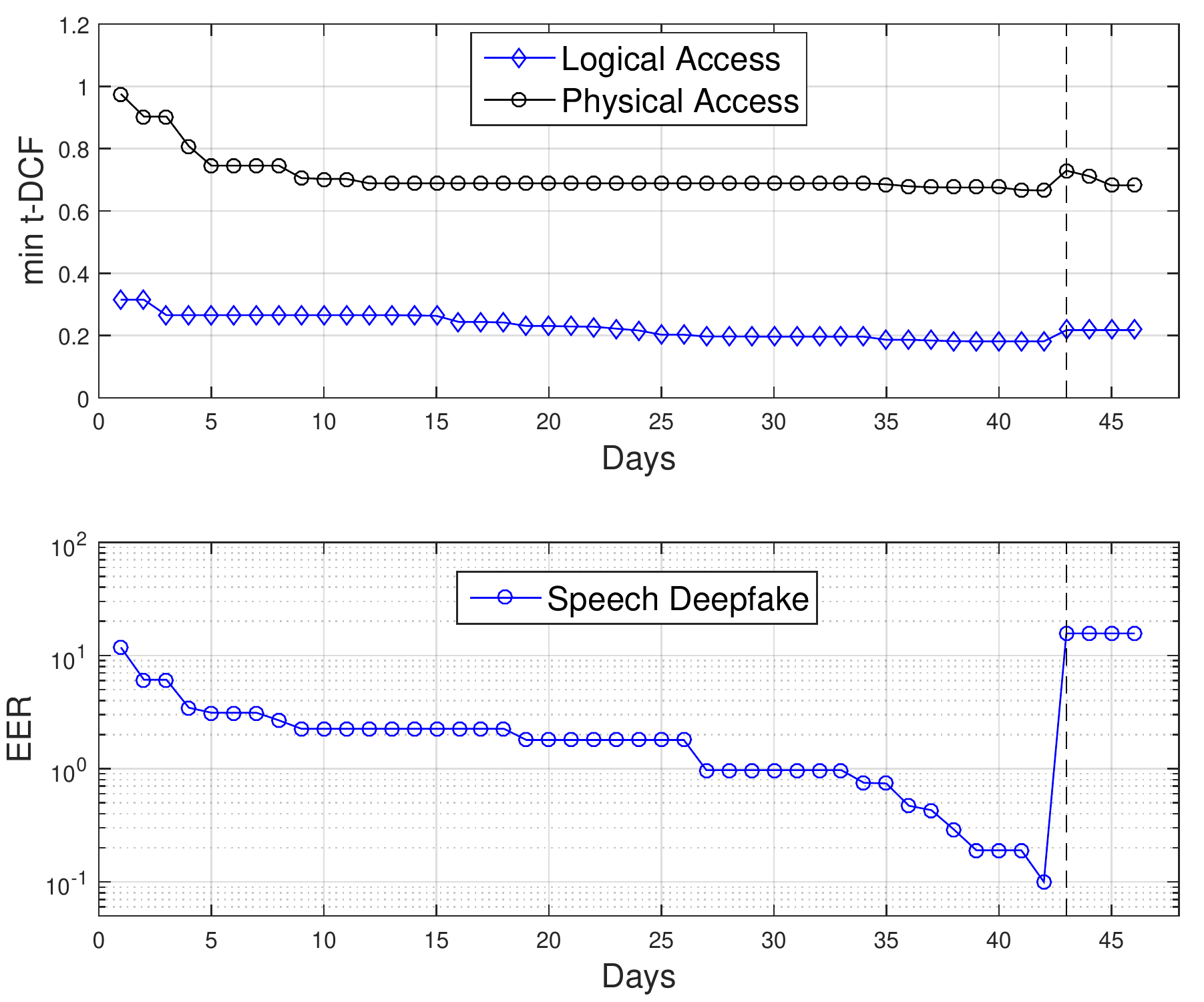}
\vspace{-3mm}
\caption{System performances on LA, PA, and DF tasks evolve steadily over the \emph{progress} and \emph{evaluation} phases.}
\label{fig:la_pa_df_progress_eval}
\end{figure}

\section{Evaluation platform}

New to the forth edition is the adoption of a web-based challenge platform.  ASVspoof 2021 used the CodaLab website through which participants could submit CM scores and receive results.  The challenge was run in two phases, with an additional post-evaluation phase (not addressed in this paper).  During the first \emph{progress} phase, which ran for six weeks, participants could submit up to three submissions per day.  Results determined from a subset of trials (the progress subset) were made available to participants who could opt to submit their results to an anonymised leaderboard.  The evaluation phase ran for only four days during which participants could make only a single submission.  This submission was evaluated using the remaining trials in the ASVspoof 2021 databases (those not used for the progress phase). 

Fig.~\ref{fig:la_pa_df_progress_eval_count} illustrates a \emph{stacked} bar chart showing the number of submissions made during the progress and evaluation phases for each of the three tasks.  Submissions increased steadily over the six-week progress phase, with fewer submissions over the weekends.  The number of submissions in week 6 is twice that in week 3 and there were more submissions for the LA task than for the PA and DF task.

\section{Challenge results}

ASVspoof 2021 adopts two different performance metrics, namely, the min t-DCF for the LA and PA tasks and EER for the DF task. Performances of all submissions over the \emph{progress} and \emph{evaluation} phases are shown in Fig. ~\ref{fig:la_pa_df_progress_eval}. Notably, the performance on PA task reached a plateau after two weeks into the \emph{progress} phase, while considerable improvement could still be observed on both LA and DF tasks closer to the end. During the \emph{progress} phase, the reduction in min t-DCF amounts to $42\%$ and $32\%$ on the LA and PA tasks, respectively. On the DF task, EER was reduced from $11.6\%$ to $0.10\%$ during the \emph{progress} phase, which proven to be illusive when the EER shoot up to $15.6\%$ in the \emph{evaluation} phase.

\begin{table}[!t]
\renewcommand{\arraystretch}{1}
\caption{ASVspoof 2021 evaluation results for the LA, PA and DF conditions.  Results shown in terms of pooled normalised minimum t-DCF and pooled EER [\%]. %
}
\label{tab:eval_results}
\scriptsize
\vspace{-5mm}
\begin{center}
\begin{tabular}{|p{.1cm}ccp{.5cm}|p{.1cm}ccp{.5cm}|}
\hline
\multicolumn{8}{|c|}{\footnotesize{\textbf{ASVspoof 2021 LA task}}} \\ \hline
\textbf{\#} & \textbf{ID} & \textbf{t-DCF} & \textbf{EER} & \textbf{\#} & \textbf{ID} & \textbf{t-DCF} & \textbf{EER} \\ \hline
1  &   T23  &  0.2177 &  1.32 & 22 &   T11  &  0.3666 &  7.19 \\ \hline
2  &   T35  &  0.2480 &  2.77 & 23 &   T34  &  0.4059 &  13.45\\ \hline
3  &   T19  &  0.2495 &  3.13 & 24 &   B04  &  0.4257 &  9.50 \\ \hline
4  &        &  0.2500 &  2.81 & 25 &   T15  &  0.4890 &  14.68\\ \hline
5  &   T36  &  0.2531 &  3.10 & 26 &   B01  &  0.4974 &  15.62\\ \hline
6  &   T20  &  0.2608 &  3.21 & 27 &   T25  &  0.5148 &  13.75\\ \hline
7  &   T08  &  0.2672 &  3.62 & 28 &   T32  &  0.5270 &  12.90\\ \hline
8  &   T16  &  0.2689 &  3.63 & 29 &        &  0.5748 &  18.50\\ \hline
9  &        &  0.2725 &  3.61 & 30 &   B02  &  0.5758 &  19.30\\ \hline
10 &   T04  &  0.2747 &  5.58 & 31 &        &  0.5775 &  14.28\\ \hline
11 &   T06  &  0.2853 &  5.66 & 32 &   T01  &  0.6204 &  15.95\\ \hline
12 &        &  0.2880 &  5.01 & 33 &        &  0.6288 &  15.87\\ \hline
13 &   T03  &  0.2882 &  4.66 & 34 &   T24  &  0.6320 &  15.98\\ \hline
14 &        &  0.2893 &  5.70 & 35 &        &  0.6371 &  16.27\\ \hline
15 &   T31  &  0.3094 &  5.46 & 36 &   T29  &  0.6741 &  17.41\\ \hline
16 &   T17  &  0.3279 &  7.19 & 37 &        &  0.6813 &  17.66\\ \hline
17 &   T07  &  0.3310 &  8.23 & 38 &   T12  &  0.7228 &  26.41\\ \hline
18 &   T30  &  0.3362 &  8.89 & 39 &        &  0.7233 &  19.19\\ \hline
19 &   B03  &  0.3445 &  9.26 & 40 &        &  0.8521 &  26.14\\ \hline
20 &   T02  &  0.3446 &  7.79 & 41 &        &  1.0000 &  53.81\\ \hline
21 &   T14  &  0.3451 &  8.98 &    &        &         &       \\ \hline
\end{tabular} \\
\end{center}
\vspace{-5mm}
\begin{center}
\begin{tabular}{|p{.1cm}ccp{.5cm}|p{.1cm}ccp{.5cm}|}
\hline
\multicolumn{8}{|c|}{\footnotesize{\textbf{ASVspoof 2021 PA task}}} \\ \hline
\textbf{\#} & \textbf{ID} & \textbf{t-DCF} & \textbf{EER} & \textbf{\#} & \textbf{ID} & \textbf{t-DCF} & \textbf{EER} \\ \hline
1  &   T07  &  0.6824 &  24.25 & 13 &        &  0.9265 &  37.10\\ \hline
2  &   T16  &  0.7122 &  27.59 & 14 &   B01  &  0.9434 &  38.07\\ \hline
3  &   T23  &  0.7296 &  26.42 & 15 &   T03  &  0.9444 &  38.07\\ \hline
4  &   T01  &  0.7446 &  28.36 & 16 &        &  0.9530 &  38.50\\ \hline
5  &   T04  &  0.7462 &  29.00 & 17 &   T09  &  0.9666 &  34.77\\ \hline
6  &        &  0.7469 &  29.22 & 18 &   B02  &  0.9724 &  39.54\\ \hline
7  &   T33  &  0.7648 &  29.55 & 19 &   T11  &  0.9939 &  42.55\\ \hline
8  &   T08  &  0.7670 &  29.02 & 20 &        &  0.9945 &  42.98\\ \hline
9  &   T37  &  0.8216 &  35.07 & 21 &   B03  &  0.9958 &  44.77\\ \hline
10 &   T27  &  0.8307 &  32.00 & 22 &   B04  &  0.9997 &  48.60\\ \hline
11 &   T26  &  0.8362 &  29.61 & 23 &   T10  &  1.0000 &  45.50\\ \hline
12 &   T28  &  0.8879 &  32.96 &    &        &         &       \\ \hline
\end{tabular} \\
\end{center}
\vspace{-5mm}
\begin{center}
\begin{tabular}{|p{.1cm}cp{.5cm}|p{.1cm}cp{.5cm}|p{.1cm}cp{.5cm}|}
\hline
\multicolumn{9}{|c|}{\footnotesize{\textbf{ASVspoof 2021 DF task}}} \\ \hline
\textbf{\#} & \textbf{ID} & \textbf{EER} & \textbf{\#} & \textbf{ID} & \textbf{EER} & \textbf{\#} & \textbf{ID} & \textbf{EER} \\ \hline
1  &  T23  &  15.64 & 12 &       &  20.63  & 23 &       &  23.88\\ \hline
2  &  T20  &  16.05 & 13 &       &  20.82  & 24 &  T35  &  24.12\\ \hline
3  &  T08  &  18.30 & 14 &  T16  &  20.84  & 25 &       &  24.89\\ \hline
4  &       &  18.80 & 15 &       &  21.61  & 26 &  T30  &  25.21\\ \hline
5  &  T06  &  19.01 & 16 &       &  21.67  & 27 &  B02  &  25.25\\ \hline
6  &  T22  &  19.22 & 17 &       &  22.03  & 28 &  T26  &  25.41\\ \hline
7  &  T03  &  19.24 & 18 &  B04  &  22.38  & 29 &  B01  &  25.56\\ \hline
8  &  T01  &  19.70 & 19 &       &  22.38  & 30 &       &  26.67\\ \hline
9  &  T36  &  20.23 & 20 &  T25  &  22.62  & 31 &  T21  &  28.96\\ \hline
10 &  T31  &  20.33 & 21 &  B03  &  23.48  & 32 &       &  29.25\\ \hline
11 &  T19  &  20.33 & 22 &       &  23.57  & 33 &       &  29.75\\ \hline
\end{tabular} 
\end{center}
\vspace{-5mm}
\end{table}

\subsection{Logical access}

Results for all four baseline systems are shown in Table~\ref{tab:eval_results} (top-most table).  The best performing B03 system gives a min t-DCF of 0.3445.  System B04 gives a min t-DCF of 0.4257.  B01 and B02 are relatively uncompetitive.

The evolution in the best performing participant submissions is illustrated in Fig.~\ref{fig:la_pa_df_progress_eval} (top plot).  Even on day one of the progress phase, the best performing system achieved a min-DCF of 0.3152, already substantially below the best performing baseline B03.  After three days, the best score dropped to 0.2656 and then, after a number of days with no improvement, it dropped gradually to 0.1815 where it remained until the progress phase ended on day~42.  

Evaluation results are shown in Table~\ref{tab:eval_results}.  The ASVspoof 2021 task attracted 42 submissions, 19 of which outperformed the best baseline B03.  The best-performing system submission made by team T23 achieved a substantial improvement beyond the best baseline, producing a min t-DCF of 0.2177, corresponding to an EER of~1.32\%.  There is a modest gap to the next-best performing system T35 from where the gap between each system is relatively small.  As observed in past editions, there are differences in system ranking when made according to min t-DCF and EER results, highlighting once again the importance of assessing the impact of spoofing and CMs in tandem with the ASV system.  Nonetheless, the submission which produced the lowest min t-DCF also achieved the lowest EER.

\subsection{Physical access}

Results for all four baseline systems are shown in Table~\ref{tab:eval_results} (middle table).  The best performing B01 system gives a min t-DCF of 0.9434, followed by B02 which gives a min t-DCF of 0.9724. B03 and B04 are less competitive, with min t-DCF closer to 1.0.

The evolution in the best performing participant submissions for the PA task is illustrated in Fig.~\ref{fig:la_pa_df_progress_eval} (top plot). During the first 10 days, the min t-DCF score of the best submission gradually decreased from around 1.0 to around 0.7. The best submission surpassed the best baseline system B01 on the 2nd day. From the 10th day to the end of the progress phase, there was no further improvement.

Evaluation phase results are shown in Table~\ref{tab:eval_results}. Among the 23 submissions, 13 outperformed the best baseline~B01. 
Although the 2021 PA task is more challenging than previous editions, the top system submitted by team T07 produced a min t-DCF of 0.68 and an EER around of 24\%, an encouraging improvement over the performance of the best baseline B01. Like for the LA task, we observe differences in ranking for min t-DCF and EER scores.

\subsection{DeepFake}

Results for all four baseline systems are shown in Table~\ref{tab:eval_results} (bottom table).  In contrast to results for LA and PA tasks, the DF task uses the EER metric.   From the four baselines, the lowest EER of 22.38\% is achieved by B04.  Baseline B03 achieved a similar score of 23.48\%,  while the other two lag further behind. %

The evolution in the best performing DF participant submissions shown in Fig.~\ref{fig:la_pa_df_progress_eval} (bottom plot) shows an improving trend during the progress phase. After an initial drop in the EER during the first ten days, the best EER decreased in piecewise fashion until day 33. Beyond this, in the remaining ~9 days before the beginning of the evaluation phase, improvements were rapid. In line with the number of submissions increasing towards the end of progress phase (Fig.~\ref{fig:la_pa_df_progress_eval_count}), this may reflect a deadline-driven approach to attempt more, or higher-risk ideas to decrease the EER. 

Evaluation results for participant submissions are shown in Table~\ref{tab:eval_results}.  From the 30 submissions, 18 yielded lower EERs than the best performing baseline. 
The best performing system submitted by team T23 produced an EER of 15.64\%, followed closely by the system submitted by team T20 which achieved an EER of 16.05\%.  Beyond, the EER increases only gradually.
Unlike for the LA and PA tasks, evaluation phase results are substantially higher than those for the progress phase.  This is an indication of overfitting to the progress partition.  While details will be revealed later, the DF evaluation set contains a number of spoofing attacks, data conditions and compression methods that differ from those seen in the progress partition.

\subsection{Statistical significance}

\begin{figure}[!h]
\centering
      \begin{subfigure}[b]{0.45\textwidth}
         \centering
         \includegraphics[width=0.75\textwidth]{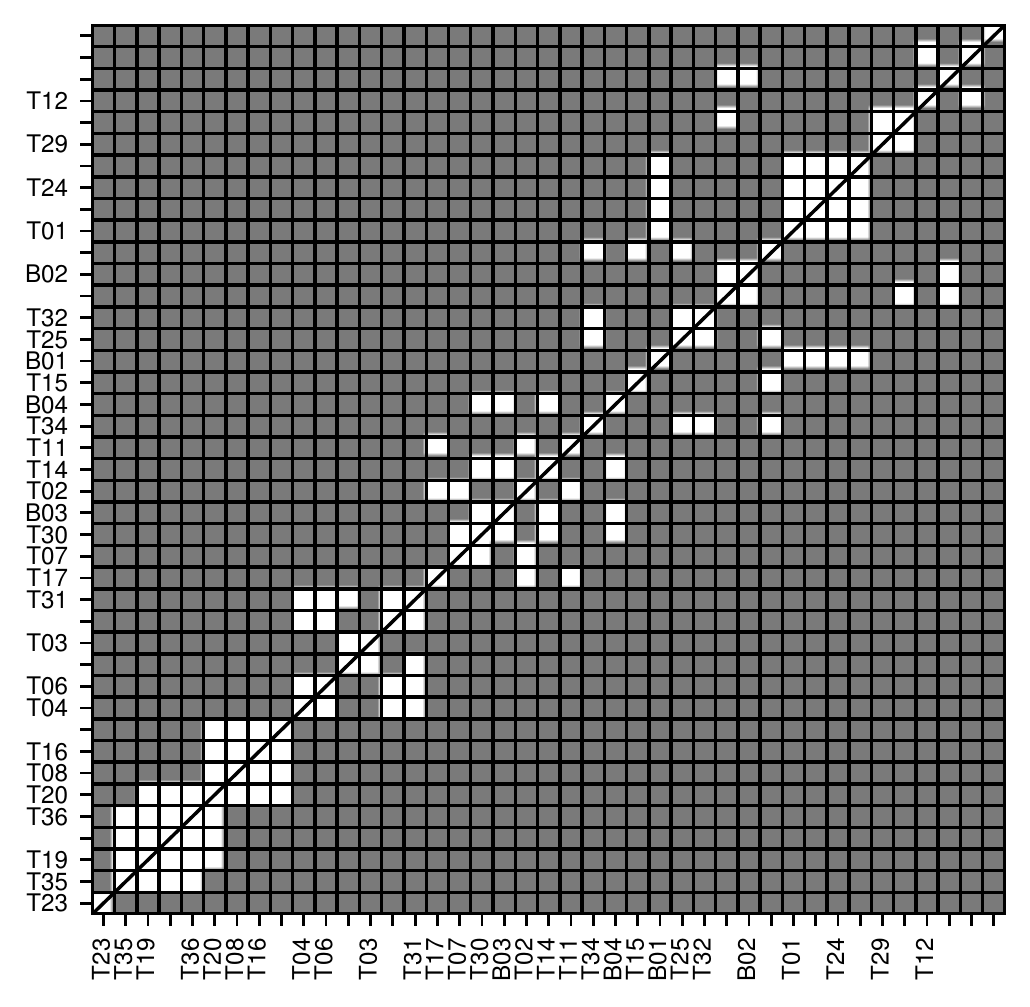}
         \vspace{-3mm}
         \caption{LA task}
         \label{fig:sigtest_LA}
     \end{subfigure}
     \begin{subfigure}[b]{0.45\textwidth}
         \centering
         \includegraphics[width=0.75\textwidth]{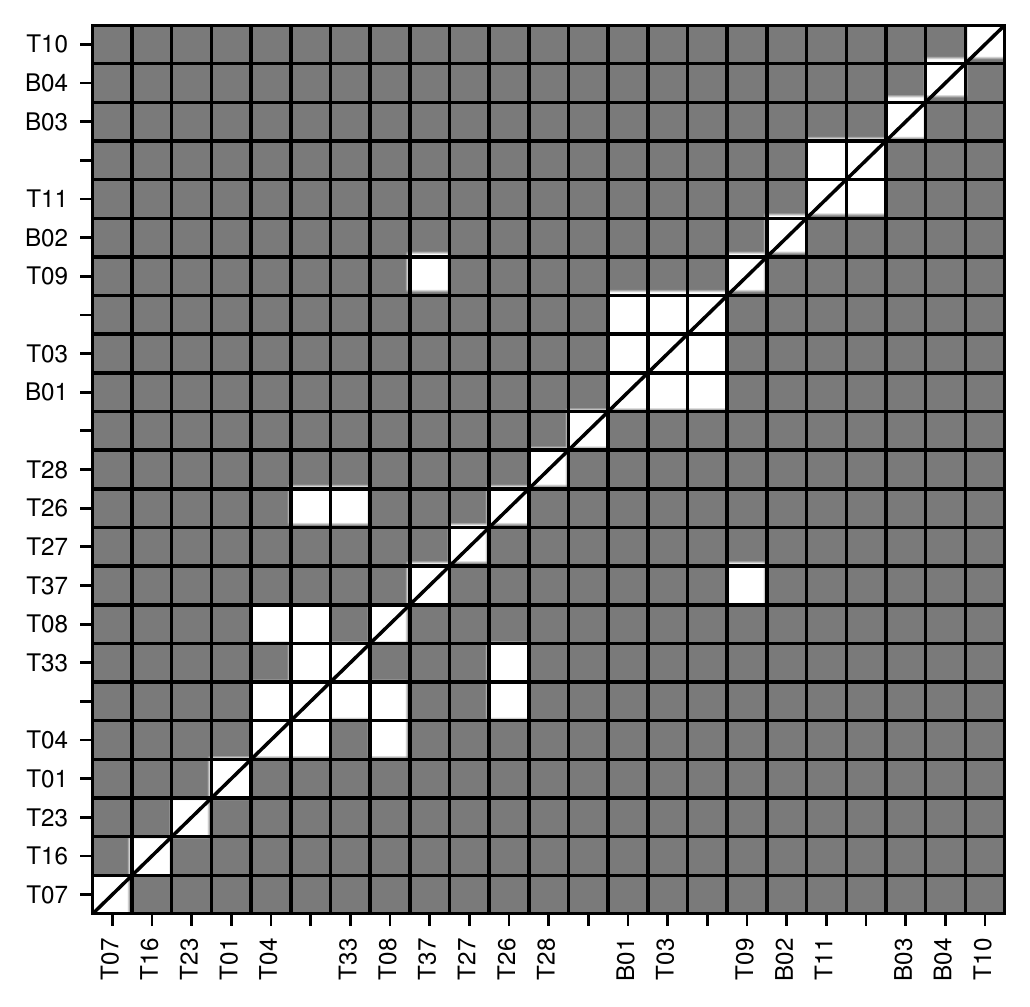}
         \vspace{-3mm}
         \caption{PA task}
         \label{fig:sigtest_PA}
     \end{subfigure}
     \begin{subfigure}[b]{0.45\textwidth}
         \centering
         \includegraphics[width=0.75\textwidth]{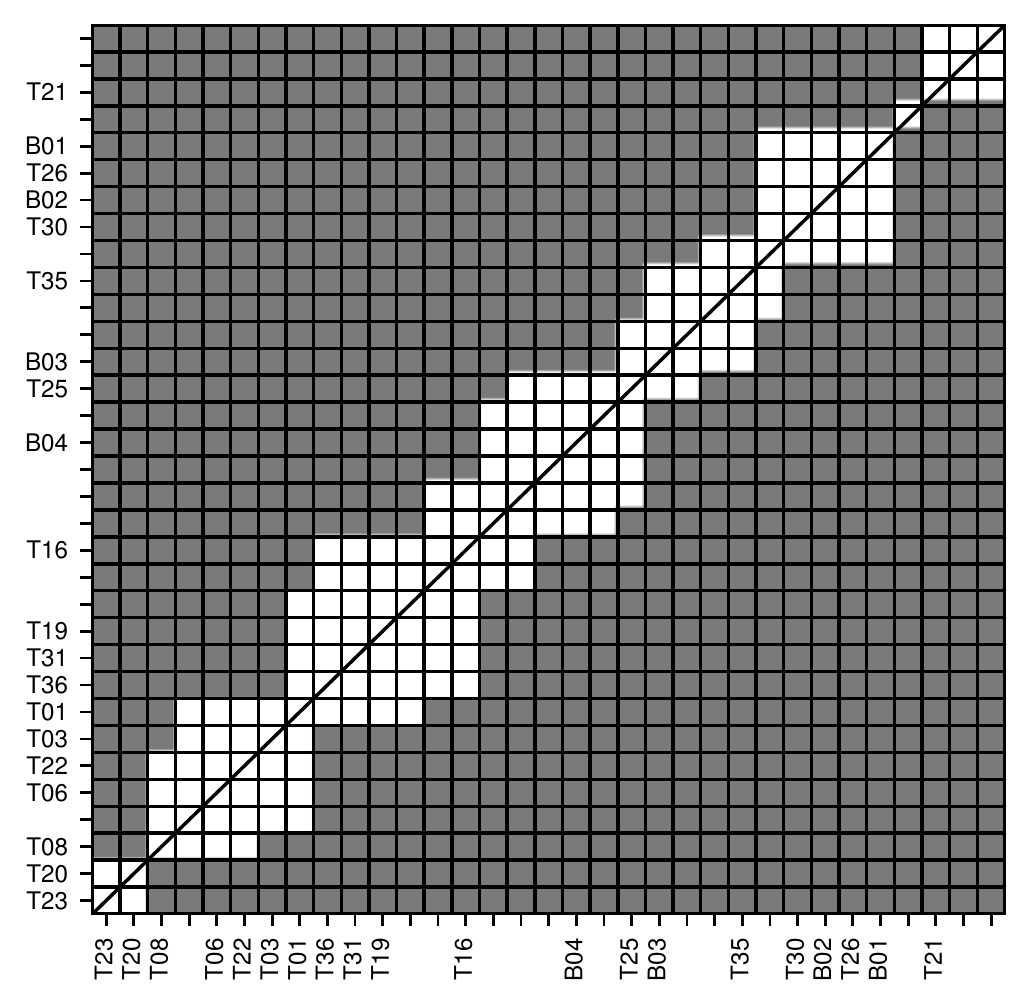}
         \vspace{-3mm}
         \caption{DF task}
         \label{fig:sigtest_DF}
     \end{subfigure}
\caption{
Pairwise statistical significance test on EERs \cite{bengio2004statistical}, given $\alpha=0.05$ with Holm-Bonferroni correction.  Significant differences are indicated in dark grey, otherwise in white.}
\label{fig:sigtest_all}
\end{figure}

Plots of statistical significance are shown in Fig.~\ref{fig:sigtest_all} for each of three tasks.  Estimated according to the approach described in~\cite{bengio2004statistical}, they show whether or not the difference in performance between any pair of systems is statistically significant.  Pairs for which performance is statistically significant are indicated by dark-grey boxes, and white otherwise.

The plots show that, for the LA and PA tasks, the performance of top performing systems is statistically significant.  For the LA task, differences in the performance of the next 8-ranked systems are less significant.  Other lower-ranked systems also show similarities to better-ranked systems.  For the PA task, differences in the performance of the top four systems are statistically significant and, again, there are similarities between the performance of lower and higher ranked systems.  For the DF task, there are 4 groups of systems. Contrary to the LA and PA conditions, the performance differences for the top two systems is not significant.

Given the complexity of the ASVspoof 2021 challenge, the addition of one new task and the extremely tight evaluation and paper submission schedule, further ongoing analysis will be reported at the ASVspoof workshop.

\section{Conclusion}

This paper outlines the ASVspoof 2021 challenge which comprised three separate tasks: logical access; physical access; speech deepfake.  The challenge was to develop spoofed/deepfake speech detection solutions.  The 2021 edition was considerably more complex than its predecessors, including not only one new task, but also more challenging data that moves ASVspoof nearer to more practical applications scenarios.  Despite the added difficulty, and despite there being no new, matched training or development data, the lowest scores for the LA task show a min t-DCF of 0.2177 and an EER of 1.32\%.  The PA and DF tasks proved more challenging, probably for different reasons.  For the PA task, the small difference in performance for progress and evaluation phases reflect the difficulty in detecting replay attacks across multiple rooms when both bonafide and spoofed data exhibit reverberation and additive noise.  For the DF task, the gap between promising progress results and notably worse evaluation results suggest a high degree of overfitting.  With the transition to an INTERSPEECH satellite event and a tighter schedule for the 2021 edition, more detailed evaluation analysis will be reported at the ASvspoof 2021 workshop and in further work. 

\section{Acknowledgements}

The ASVspoof 2021 organising committee extends its sincere gratitude and appreciation to the 72 participants and teams.
For reasons of anonymity, they could not be identified in this article.  Subject to the publication of their results and prior approval, they will be cited or otherwise acknowledged in future work.

\bibliographystyle{IEEEbib}

\clearpage
\balance

\bibliography{main}

\end{document}